\documentclass[12pt]{article}
\usepackage{bezier}
\usepackage{amssymb}
\usepackage{epsfig}

\title{
Catastrophic senescence of the Pacific salmon without mutation-accumulation
}
\author{Hildegard~Meyer-Ortmanns\\
Institut f\"ur Theoretische Physik\\
Universit\"at Heidelberg \\
Philosophenweg 16\\
D-69120 Heidelberg, Germany \\
e-mail: ortmanns@thphys.uni-heidelberg.de}

\begin{document}

\maketitle

\begin{abstract}
\setlength{\baselineskip}{1pt}
\noindent
We derive catastrophic senescence of the Pacific salmon from an aging 
model which
was recently proposed by Stauffer. The model is based on the postulates
of a minimum reproduction age and a maximal genetic lifespan. It allows
for self-organization of a typical age of first reproduction and a typical 
age of death. Our Monte Carlo
simulations of the population dynamics show that the model leads to 
catastrophic senescence for semelparous reproduction as it 
occurs in the case of 
salmon, but to a more gradually increase of senescence for iteroparous
reproduction.\\
\vskip5pt
\noindent{\bf Keywords}: population dynamics, aging, Stauffer model, Monte Carlo
\end{abstract}


\section{Introduction}

According to the evolutionary theory of aging senescence is ultimately
caused by a declining pressure of natural selection as one gets older and
older \cite{rose}.
One manifestation of senescence (which can be most easily handled on
the computer) is an increase of mortality with age or, more precisely,
a decrease of 
survivors at age $a$, from an initial population of newly born 
offspring at time $0$. Senescence is attributed to several factors like
environment, metabolism, and, most important, by several genetic mechanisms.
Two major genetic mechanisms, which are under discussion and need not 
necessarily exclude each other, are antagonistic pleitropy and mutation-
accumulation. Based
on the latter mechanism, i.e. on the hypothesis of an increase of deleterious
mutations with age, Penna et al. (1995)
proposed a bit-string
model \cite{penna} , 
which nowadays is widely used for Monte Carlo simulations of
aging, because it predicts many experimentally observed features
of senescence (for a recent review see \cite{war}). 
One of the successful predicitions is the catastrophic
senescence for the Pacific salmon. 

Pacific salmon show the most dramatic manifestation of aging. As semelparous
individuals they breed only once in their life, all at the same age, with
plenty of offspring, and die a few weeks later 
. In contrast to salmon, 
iteroparous individuals breed repeatedly and age more gradually.

\section{The model}

Recently Stauffer has proposed a model which is based on the 
postulate of a minimum reproduction age and a maximal genetic lifespan
\cite{stauffer}. 
Only these two numbers are transmitted from generation to generation, 
with certain mutations, by asexual reproduction. The population 
consists of $N$ individuals $i$
$(i \in 1,...,N)$ initially. Each individual is characterized by three 
integers: its age $a(i)$, its minimum reproduction age $a_m(i)$ and its
maximal genetic lifespan $a_d(i)$ with $0\leq a_m(i)<a_d(i)\leq 32$.
The maximal lifetime is
restricted to 32 time units (called years), the  
minimum reproduction age may be chosen between zero and $a_d(i)-1$. 
Within these constraints the values
of $a_m(i)$ and $a_d(i)$ are randomly mutated for an offspring by $\pm1$,
away from the maternal values, and for each child separately. These mutations
realize some kind of antagonistic pleiotropy in the sense that a shorter 
lifespan or a later reproduction age increase the birth rate. (``Parents die
to make place for their children''.) After an individual 
has reached its minimum reproduction age, it gives birth to one 
offspring with probability $b$, chosen as 
$(1+\epsilon)/(a_d(i)-a_m(i)+ \epsilon)$ with $\epsilon=0.08$ for convenience. 
Independently of the genetic death, 
which happens automatically
and unavoidably if $a(i)=a_d(i)$, at each time interval an individual can 
also die ``accidentally'', with the Verhulst probalility $N/N_{max}$. $N_{max}$
is called the carrying capacity to account for the fact that any given 
environment can only support populations up to some maximal size $N_{max}$. 
Otherwise the individuals die because of food and space limitations. Stauffer's
model shows the basic features required for senescence. The age distribution
shows an increase of mortality with age. Moreover, a self-organization of a 
typical age of first reproduction and of death is observed, 
similarly to Ito's self-organization
of a minimum reproduction age in the framework of the Penna model \cite{ito}.
\vskip7pt
In this paper we apply Stauffer's model to the population dynamics of Pacific
salmon and check whether the model is able to reproduce the qualitative 
features of catastrophic senescence. For a given value of $N_{max}$
and an initial population of $N$ individuals we specify $a_m(i)=1$
and $a_d(i)=16$ for all $i=1,...,N$ as initial values. One Monte Carlo
iteration then consists of the following three steps that we call ``deaths'',
``births'' and ``aging''.
\begin{itemize}
\item
In the first loop of ``deaths'' over all individuals, each individual dies
either with the Verhulst probability $N/N_{max}$, or, if it survives space and 
food limitations, because it has reached its maximal genetically allowed age
$a_d(i)$. Otherwise it survives. The initial population of size $N$
gets reduced this way.
\item
In the second loop of ``births'' over all individuals, each individual gives 
birth to $n_b\geq 1$ offspring with probability $b=
1.08/(a_d(i)-a_m(i)+ 0.08)$, provided the maternal age is not below the 
minimum reproduction age and equals a fixed given integer $a_0$
with $0\leq a_m(i)\leq a_0<a_d(i)\leq 32$ and $a_0(i)=a_0$ is chosen
the same for
all individuals. The latter condition obviously accounts for the specific
features of salmon which breed only once and all at the same age.
It turns out that the choice of the number of offspring $n_b$ is not arbitrary,
because the second condition is so restrictive that the population only 
survives for sufficiently large $n_b$. Now the values of
$a_m$ and $a_d$ for the offspring are mutated away from the maternal values,
again at each time interval by $\pm1$ for each child separately as in 
Stauffer's model, but with probability $p_s<1$. This way the mutations become
suppressed with probability $1-p_s$, to account for the experimental 
fact that the
time interval for reproduction of salmon is rather narrow for {\it any} 
generation. Unless mutations are suppressed, they have the tendency of
spreading the first reproduction age within a broader time interval. On the
histograms of aging this has a similar effect as iteroparous reproduction.
The question is whether the minimum reproduction age adjusts itself to a value
which is self-consistent with the prescribed fixed reproduction age $a_0$.
The remaining mutations should drive the initial values for $a_m(i)$ towards
$a_0$.
\item
In the third step of a single iteration, the population which remains from
the first two steps ages by one time unit, $(a(i)\to a(i)+1)$. The population
size and averages over the individual minimum reproduction ages and the
maximal lifespans are stored. From a certain number of iterations on
also the individual ages $a(i)$, $a_m(i)$ and $a_d(i)$ are stored in 
histograms as a function of the time interval $j$, $j\in (1,...,32)$.
\end{itemize}

Now the iterations are repeated a number of $t$ times until the population
dynamics has stabilized and the fluctuations in average values are negligibly
small.

\section{Results}
Figs.1 and 2 show various histograms of ages $a_d$
for genetic death which were obtained for the following choice of parameters.
In Fig.1 the carrying capacity $N_{max}$ is chosen as $2\cdot 10^5$,
$N$ initially as $N_{max}/2$. The number of iterations $t$ is 
$2\cdot 10^4$. The
actual reproduction age $a_0$ is fixed to 6, and the number of births $n_b$
which an individual can give to offspring at the age of 6 is 12. Mutations
of the minimum reproduction age of offspring are suppressed by $80\% (+)$,
$95\% (\mbox{x})$ and $99\% (\star)$, respectively. The histograms show a clear
self-organized maximum of genetic death at an age of 7. In discrete time
units this means that death occurs most likely directly after reproduction.
The peaks are the sharper the stronger the suppression of mutations. The
qualitative shape of the histograms stays the same when $N_{max}$ is varied
over several orders of magnitude. Already after $\approx 100$ iterations
the population dynamics stabilizes in the sense that the population size 
oscillates regularly between stable minimum and maximum values. These regular
oscillations build up randomly during the first $100$ iterations and
then are reproduced with the period $a_0=6$.
The value of $n_b=12$ is the
minimal integer so that the population with $95\%$ suppression of mutations
survives as a whole. For smaller $n_b$ it dies out after a few iterations.
This feature is in qualitative agreement with nature. Pacific salmon produce
plenty offspring once they breed to compensate for the restrictive conditions
on $a_m$ and $a_0$ independent of $i$.
The maximum of the histogram of ages for the minimal reproduction age lies 
at an age of 6, consistent with the prescribed value $a_0$.

In Fig.2 we compare histograms of ages for genetic death between
iteroparous (full line) and semelparous (dashed line) reproduction. The
dashed curve was obtained for $N_{max}=2\cdot 10^5$, $t=2\cdot 10^4$, 
$a_0=6$, $n_b=12$, $1-p_s=99\%$.
Again it shows a sharp peak at the age of 7. The histogram for 
iteroparous reproduction was obtained for $N_{max}=2\cdot 10^7$, 
$t=2\cdot 10^4$, reproduction can happen at 
any age between $a_m(i)$ and $a_d(i)$, (i.e. i-dependent and
possibly several times in one life), 
$n_b=1$ and no suppression of mutations of 
$a_m(i)$ and $a_d(i)$ for the offspring. The hundred times larger value for
$N_{max}$ was chosen for convenience to get comparable numbers for $a_d$ in
spite of the different reproduction conditions. The maxima of both curves
are self-organized, but in the iteroparous case the maximum is much broader,
going along with a gradual increase of senescence with time or a less rapid
aging than in the case of salmon, compatible with the Penna model \cite{penna}.

\section{Summary}

Stauffer's model based on the two postulates of a minimum reproduction age
and a maximal genetic lifespan predicts the catastrophic senescence
for the Pacific salmon in a qualitative way. The larger the suppression of 
mutations, the faster die the individuals after reproduction, and the larger
is the number of births in one reproduction step which is necessary to sustain
the species. The self-organized maximum of the histogram of ages $a_m$ is
self-consistent with the only allowed age $a_0$ for reproduction, a restriction
that is used as an input. Stauffer's explanation works 
without mutation-accumulation, whereas 
mutation-accumulation was
an essential ingredient in the Penna model. Therefore,
differently from what the success of the Penna model might have suggested, 
mutation-accumulation does not seem to be essential for 
explaining gross qualitative features of senescence.

\section{Acknowlegdment}

I would like to thank Dietrich Stauffer for stimulating discussions.

\newpage

\begin{center}
\begin{figure}[hb]
\mbox{\epsfig{file=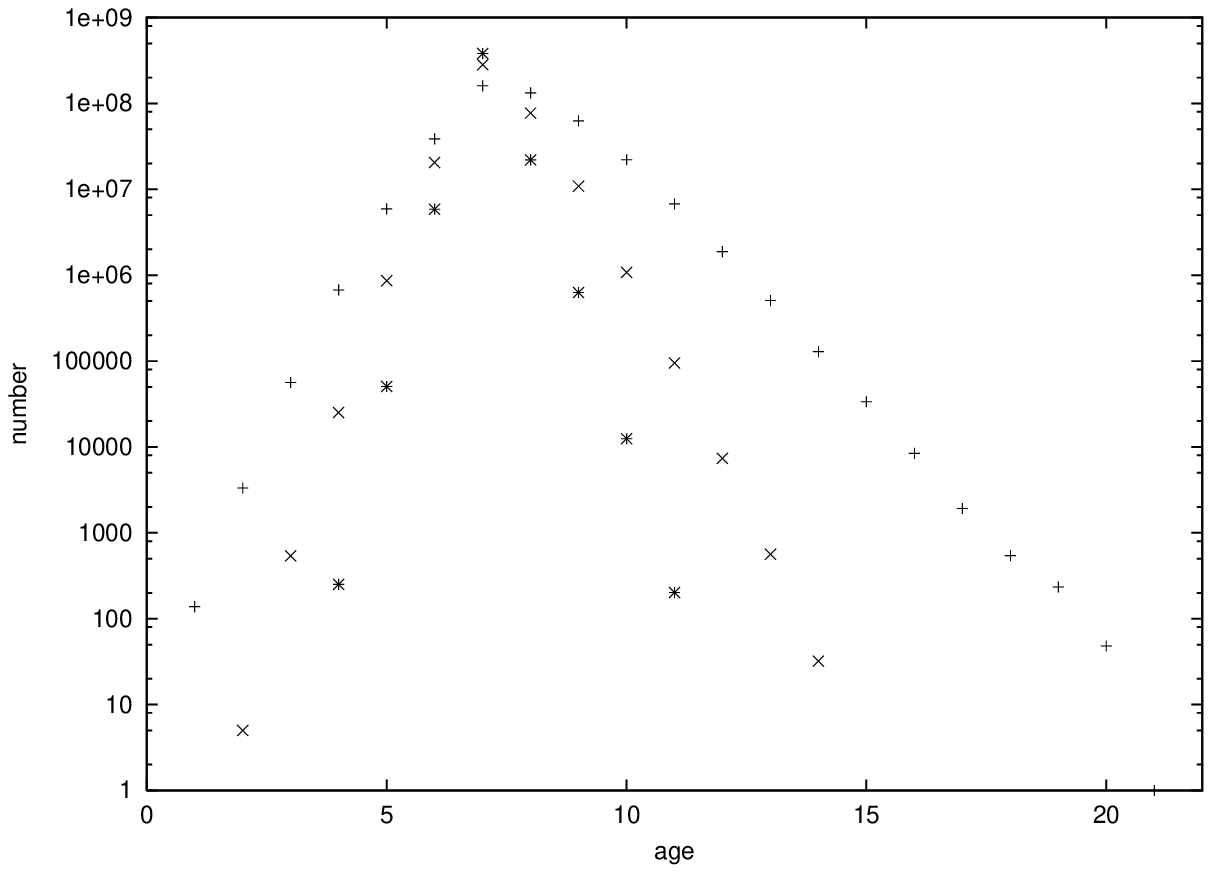,width=15cm}}
\caption{Histogram of ages for genetic death; mutation probability $20(+)$,
$5(\mbox{x})$ and $1(\star)$ percent}
\end{figure}
\end{center}    
\begin{center}
\begin{figure}[hb]
\mbox{\epsfig{file=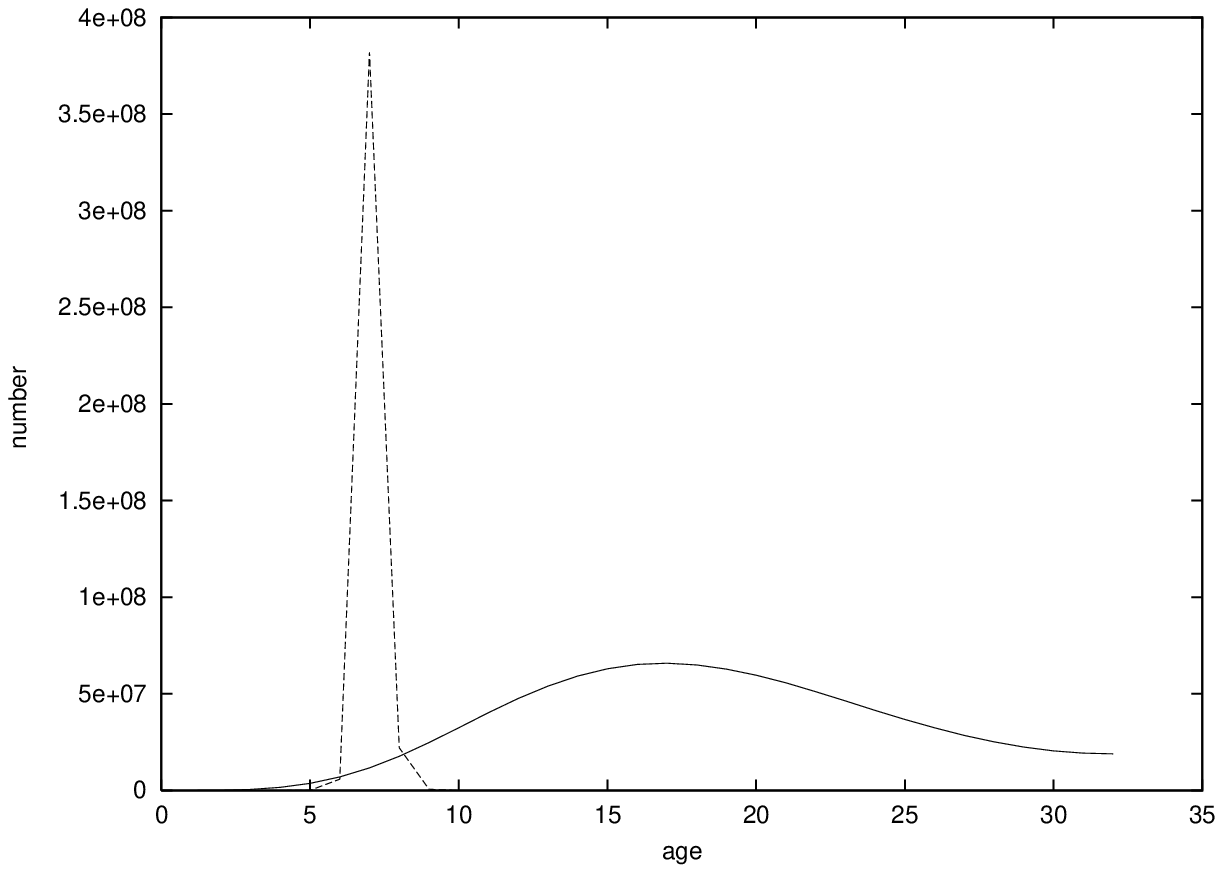,width=15cm}}
\caption{Histogram of ages for genetic death: iteroparous (full) and 
semelparous
(dashed) reproduction}
\end{figure}
\end{center}       


\begin{thebibliography}{9}
\bibitem{rose} M.~R.~Rose, Evolutionary Biology of Aging (Oxford University 
Press, New York, 1991).
\bibitem{penna} T.~J.~P.~Penna, S.~Moss de Oliveira and D.~Stauffer, Phys.~Rev.
~E 52, R3309 (1995).
\bibitem{war} S.~Moss de Oliveira, P.~M.~C.~de Oliveira and D.~Stauffer,
Evolution, Money, War and Computers (Teubner, Stuttgart-Leipzig 1999).
\bibitem{stauffer} D.~Stauffer, in Biological Evolution and Statistical Physics
(Dresden, May 2000), M.~L\"assig and A.~Valleriani (Springer, Berlin 2001 or 
2002).
\bibitem{ito} N.~Ito, Int.~J.~Mod.~Phys.~C 7, 107 (1996).
\end{thebibliography}
\end{document}